\documentclass[usenatbib]{mn2e}
\usepackage{graphicx,rotate}
\title{Dynamical friction force exerted on spherical bodies}
\author[O. Esquivel \& B. Fuchs]{O.~Esquivel\thanks{Fellow of IMPRS for
Astronomy and Cosmic Physics, Heidelberg}
and B.~Fuchs\\
Astronomisches Rechen-Institut am Zentrum f\"ur Astronomie der 
Universit\"at Heidelberg,\\
M\"onchhofstra{\ss}e 12-14, 69120 Heidelberg, Germany}

\begin{document}

\maketitle

\begin{abstract}
We present a rigorous calculation of the dynamical friction force exerted on a
spherical massive perturber moving through an infinite homogenous system of
field stars. By calculating the shape and mass of the polarization cloud 
induced by the perturber in
the background system, which decelerates the motion of the perturber, we 
recover Chandrasekhar's drag force law with a modified Coulomb logarithm. As 
concrete examples we calculate the drag force exerted on a Plummer sphere or 
a sphere with the density distribution of a Hernquist profile. It is shown 
that the shape of the perturber affects only the exact form of the Coulomb
logarithm. The latter converges on small scales, because encounters of the test
and field stars with impact parameters less than the size of the massive 
perturber become inefficient. We confirm this way earlier results based on the
impulse approximation of small angle scatterings. 
\end{abstract}

\section{Introduction}

The determination of the dynamical friction force exerted by a gravitating
system of background stars on a test star moving through the system is one of 
the classical problems of stellar dynamics. In his seminal paper \citet{C43}
envisaged the scenario of a sequence of consecutive gravitational two--body
encounters of test and field stars in order to calculate the drag force
(cf.~\citealt{H73} for a modern presentation). In an alternative approach 
\citet{M68} and \citet{K72} determined the collective response of the 
background medium to a moving massive perturber which manifests itself as a
polarization cloud decelerating the perturber. The dynamical friction force law 
derived in this way is precisely identical to Chandrasekhar's force law, if the
mass of an individual field star can be neglected against the mass of the 
perturber. The latter is to be expected whenever the background can be treated 
as a gravitating continuum.

The applications of Chandrasekhar's formula covered a wide field, ranging
from calculations of the dynamical friction and diffusion coefficients of the
Fokker-Planck equation, which is used to  describe the internal dynamics of
star clusters, to determinations of the sinking rates of satellite galaxies 
which are accreted by massive galaxies. The literature on these subjects has
become so extensive over the years that we do not attempt any detailed review 
here, because this would be beyond the scope of this paper. In the context of 
the sinking satellite problem most of the efforts concentrated on adapting 
the original concept of an infinite homogeneous background of field stars on
isotropic straight--line orbits to geometries and kinematics appropriate for 
the discs and haloes of spiral galaxies. Moreover, the finite size of the
perturbing body was taken into account. \citet{W76} followed Chandrasekhar's 
concept of a sequence of two--body encounters of field and test stars and
considered a sequence of deflections of field stars by a moving spherical body. 
Technically the deflections were calculated using the impulse
approximation of small angle scatterrings. The principal result was a 
modification of the Coulomb logarithm so that it does not diverge anymore at
small scales, because gravitational encounters at impact parameters smaller 
than the size of the perturbing body become ineffective.

In this paper we present a rigorous calculation of the dynamical friction force 
exerted by an infinite homogenous background on a spherical massive body
using the wave--mechanical method of \citet{M68} and \citet{K72}. This approach
is topically closely related to studies of the exchange of angular and linear
momentum in stellar systems (\citealt{LBK72}, \citealt{D76}, \citealt{TW84},
\citealt{F04}) or in plasmas (\citealt{ST92}).
It is shown in section 4 below that the shape of the perturber affects only 
the exact form of the Coulomb logarithm. As concrete examples we calculate the
drag force exerted on a Plummer sphere and on a sphere with the density
distribution of a \citet{H90} profile, respectively, and compare this with the
drag force exerted on a point mass.

\section{Induced polarization clouds}
 
We assume an infinite homogenous distribution of field stars on isotropic 
straight--line orbits. The response of the system of background stars to the
perturbation due to a massive perturber is determined by solving the
linearized Boltzmann equation
\begin{equation}
\frac{\partial f_1}{\partial t} + \sum_{i=1}^3 {\rm v}_{\rm i} 
\frac{\partial f_1}{\partial {\rm x}_{\rm i}}  - 
\frac{\partial \Phi_1}{\partial {\rm x}_{\rm i}}
\frac{\partial f_0}{\partial {\rm v}_{\rm i}} = 0\,,
\label{eq1}
\end{equation}
where $\Phi_1$ denotes the gravitational potential of the perturber and 
$f_1$ is the induced perturbation of the distribution function of the field 
stars in phase space. The unperturbed distribution function is described 
by $f_0$. The solution of the Boltzmann equation is greatly facilitated by 
considering Fourier transforms of the perturbations of the distribution 
function and the potential, respectively,
\begin{equation}
f_{\omega,{\bf k}};\, \Phi_{\omega,{\bf k}} \, \rmn{exp}i[\omega t + 
{\bf k}\cdot{\bf x}]\,,
\label{eq2}
\end{equation}
where $\omega$ and ${\bf k}$ denote the frequency and wave vector of the 
Fourier  components. Without loss of generality the spatial coordinates 
${\rm x}_{\rm i}$ and the corresponding velocity components ${\rm v}_{\rm i}$
can be oriented with one axis parallel to the direction of the wave vector.
The Boltzmann equation (\ref{eq1}) takes then the form
\begin{equation}
\omega f_{\omega,{\bf k}} + \upsilon kf_{\omega,{\bf k}} - 
k\Phi_{\omega,{\bf k}}\frac{\partial f_0}{\partial \upsilon} = 0
\label{eq3}
\end{equation}
with $k=|{\bf k}|$ and $\upsilon$ denoting the velocity component parallel
to ${\bf k}$. Equation (\ref{eq3}) has been integrated over the two velocity
components perpendicular to ${\bf k}$. In the following we assume for the
field stars always a Gaussian velocity distribution function, or specifically
in eq.~(\ref{eq3})
\begin{equation}
\frac{\partial f_0}{\partial \upsilon} = - \frac{\upsilon}{\sigma^2}
\frac{n_{\rm b}}{\sqrt{2\pi}\sigma}e^{-\frac{\upsilon^2}{2\sigma^2}}\,,
\label{eq4}
\end{equation}
where $n_{\rm b}$ denotes the spatial density of the field stars. We find then
the solution of the Fourier transformed Boltzmann equation
\begin{equation}
f_{\omega,{\bf k}}=-\frac{kv}{\omega+k\upsilon}
\frac{n_{\rm b}}{\sqrt{2\pi}\sigma^3}e^{-\frac{\upsilon^2}{2\sigma^2}} 
\Phi_{\omega,{\bf k}}\,.
\label{eq5}
\end{equation}

Integrating eq.~(\ref{eq5}) over the $\upsilon$--velocity leads to the 
density distribution of the induced polarization cloud. This has been calculated
here without taking into account the self--gravity of the background medium. 
\cite{F04} has shown that in linear approximation the effects of 
self--gravity can be described by another linearized Boltzmann equation of 
the form of eq.~(\ref{eq1}) where $\Phi_1$ denotes then the gravitational
potential of the density perturbation of the background medium. In a 
self--gravitating system the density perturbations are the sources of the 
potential perturbations so that the density - potential pair has to fulfill 
the Poisson equation. The solution
of the combined Boltzmann and Poisson equations describes simply the Jeans 
collapse of the background medium on scales larger than the Jeans length. In
real stellar systems or dark haloes their Jeans length will be always larger
than the size of the system, because otherwise the system would have collapsed 
to smaller sizes. Since the polarization cloud is contained within the system,
self--gravity is not important for its dynamics.

Of course
real self--gravitating systems are not homogeneous, but their density falls 
off radially. Moreover, their phase space distribution is more complicated 
than an isothermal Gaussian velocity distribution. In such systems a moving 
perturber can incite large scale perturbations which contribute to the 
dynamical friction of the perturber as well (\citealt{TW84},
\citealt{Cetal99}). Obviously these effects cannot be described by the 
simplified model adopted here. However, \cite{Cetal99} among others have shown,
that the drag by the localized polarization cloud on the moving perturber, 
which we treat here, is the principal effect.

\section{Potentials of the perturbing bodies}

For reference reasons we include in our analysis also the potential of a
point mass, which moves with the velocity ${\rm v}_0$ along the y--axis,
\begin{equation}
\Phi_1=-\frac{Gm}{\sqrt{{\rm x}^2 + ({\rm y}-{\rm v}_0 t)^2 + {\rm z}^2}}\,.
\label{eq6}
\end{equation}
Its Fourier-Transform can be calculated using formulae (3.754) and (6.561) of
\citet{GR00} as
\begin{equation}
\Phi_{\bf k}=-\frac{Gm}{2\pi^2}\frac{1}{k^2}e^{-i{\rm k}_{\rm y}{\rm v}_0 t}\,.
\label{eq7}
\end{equation}
Next, the potential (\ref{eq6}) is generalized to
\begin{equation}
\Phi_1=-\frac{Gm}{\sqrt{r_0^2 + {\rm x}^2 + ({\rm y}-{\rm v}_0 t)^2 + 
{\rm z}^2}}\,,
\label{eq8}
\end{equation}
which corresponds to an extended body with the mass distribution of a Plummer
sphere (\citealt{BT87}),
\begin{equation}
\rho =\frac{m}
{\frac{4\pi}{3}r_0^3}\left(1+\frac{r^2}{r_0^2}\right)^{-\frac{5}{2}}\,.
\label{eq8a}
\end{equation}
 
The Fourier transform of a moving Plummer sphere is given by
\begin{equation}
\Phi_{\bf k}=-\frac{Gm}{2\pi^2}\frac{r_0}{k}K_1(kr_0)
e^{-i{\rm k}_{\rm y}{\rm v}_0 t}\,,
\label{eq9}
\end{equation}
where $K_1$ denotes the modified Bessel function of the second kind.
As third example we consider a perturber which has the mass density 
distribution of a \citet{H90} profile,
\begin{equation} 
\rho = \frac{mr_0}{2 \pi}\frac{1}{ r(r_0+r)^3}.
\label{eq9a}
\end{equation}
Its gravitational potential is given by
\begin{equation}
\Phi_1 = - \frac{Gm}{r_0+r}\,.
\label{eq10}
\end{equation}
The Fourier transform of a moving Hernquist sphere can be calculated using
eq.~(3.722) of \citet{GR00}\footnote{We use the identity 
$\rmn{sin}kr=-\frac{1}{r}\frac{\partial}{\partial k}\rmn{cos}kr$ 
in eq.~(3.722).}
leading to
\begin{eqnarray}
\Phi_{\bf k}=-\frac{Gm}{2\pi^2}\frac{1}{k^2}
[1 + kr_0\,\rmn{cos}(kr_0)\rmn{si}(kr_0)\nonumber \\
 - kr_0\,\rmn{sin}(kr_0)\rmn{ci}(kr_0)] e^{-i{\rm k}_{\rm y}{\rm v}_0 t}\,,
\label{eq11}
\end{eqnarray}
where si and ci denote the sine- and cosine-integrals, respectively.
The model of a Plummer sphere has often been used in numerical simulations of 
the accretion and their eventual disruption of satellite galaxies in massive 
parent galaxies. Plummer spheres have constant density cores, whereas 
numerical simulations of the formation of galactic haloes in cold dark 
matter cosmology show that dark haloes may have a central density cusp 
(\citealt{NFW97}). Thus models of a Plummer or a Hernquist sphere should 
encompass the range of plausible models for satellite galaxies. The density
in both models falls off radially steeper than found in the cold dark matter 
galaxy cosmogony simulations. This mimics the tidal truncation of satellite 
galaxies in the gravitational field of their parent galaxies. A more technical 
point is that a density profile as shallow as $r^{-3}$ as suggested by 
\citet{NFW97} lends itself not easily to an analytic treatment, because the 
total mass would be diverging without an outer cut-off. 

\section{Dynamical friction}

The ensemble of stars is accelerated by the moving perturber as
\begin{equation}
<\dot{\bf v}> = - \int d^3{\rm x} \int d^3{\rm v} f({\bf x},{\bf v}) \nabla
\Phi_1 \,,
\label{eq11a}
\end{equation}
where $f$ denotes the full distribution function $f=f_0+f_1$. The 
contribution from $f_0$ cancels out, and introducing the Fourier transforms 
(\ref{eq2}) we find
\begin{eqnarray}
<\dot{\bf v}>=
-\int d^3{\rm x} \int d^3{\rm v} 
\int d^3{\rm k}\,  i {\bf k}\Phi_{\bf k}
\nonumber \\ \times  e^{i[\omega t +{\bf k}\cdot{\bf x})]} 
\int d^3{\rm k'} f_{\bf k'} 
e^{i[\omega' t + {\bf k}'\cdot{\bf x}]} \,.
\label{eq12}
\end{eqnarray}
From symmetry reasons the  acceleration vector $<\dot{\bf v}>$ is 
expected to be oriented along the y--axis.
In eq.~(\ref{eq12}) the frequency $\omega$, and
similarly $\omega'$, is given according to eqns.~(\ref{eq7}),(\ref{eq9}) and 
(\ref{eq11}) by $\omega=-{\rm k}_{\rm y} {\rm v}_0-i\lambda$ where we have
introduced following Landau's rule a negative imaginary part, which we will let 
go to zero in the following.
Moreover, $\Phi^*_{\bf k} = \Phi_{-{\bf k}}$ so that the potential is a
real quantity. Equation (\ref{eq12}) simplifies to
\begin{eqnarray}
<\dot{\bf v}>=(2\pi)^3\int d^3{\rm v}
\int d^3{\rm k}\, i{\bf k}
\Phi_{-{\bf k}} f_{\bf k} e^{2\lambda t}\,.
\label{eq13}
\end{eqnarray}
Using expression (\ref{eq5}) this can be evaluated as
\begin{eqnarray}
<\dot{\bf v}>=-\frac{(2\pi)^{5/2} n_{\rm b}}{\sigma^3}
\int d^3{\rm k}\int^{\infty}_{-\infty}d\upsilon\,
\frac{ik{\bf k}\upsilon}{-{\rm k}_{\rm y}{\rm v}_0-i\lambda+k \upsilon }
\nonumber \\
\times |\Phi_{\bf k}|^2e^{2\lambda t}e^{-\frac{\upsilon^2}{2\sigma^2}} \\
=\frac{(2\pi)^{5/2} n_{\rm b}}{\sigma^3}\int d^3{\rm k}
\int^{\infty}_{-\infty}d\upsilon \, 
\frac{k{\bf k}\upsilon\lambda}{(k\upsilon - {\rm k}_{\rm y} {\rm v}_0 )^2
+\lambda^2}\nonumber \\
\times |\Phi_{\bf k}|^2e^{2\lambda t}e^{-\frac{\upsilon^2}{2\sigma^2}}
\nonumber \,.
\label{eq14}
\end{eqnarray}
Next we observe that
\begin{equation}
\lim_{\lambda\to 0}\ e^{2\lambda t} 
\frac{\lambda}{({\rm k}_{\rm y}{\rm v}_0-k \upsilon )^2+\lambda^2}=
\pi\delta(k \upsilon -{\rm k}_{\rm y}{\rm v}_0)
\label{eq15}
\end{equation}
so that
\begin{eqnarray}
<\dot{\bf v}>=\frac{(2\pi)^{5/2} \pi n_{\rm b}}{\sigma^3}
\int d^3{\rm k}\,|\Phi_{\bf k}|^2\frac{\bf k}{k}
\, {\rm k}_{\rm y}{\rm v}_0 
e^{-\frac{({\rm k}_{\rm y}{\rm v}_0)^2}{2k^2\sigma^2}}\,.
\label{eq16}
\end{eqnarray}
The Fourier transform of any potential with spherical symmetry depends only 
on $k=|\bmath{k}|$. Thus it follows immediately from eq. (\ref{eq16}) that 
indeed the two acceleration components
\begin{equation}
<\dot{\rm v}_{\rm x}>\ =\ <\dot{\rm v}_{\rm z}>\ =\ 0
\label{eq17}
\end{equation}
as anticipated from symmetry reasons.
Only in the direction of motion of the perturber there is a net effect. 
According to Newton's third law the drag force exerted on the perturber is 
given by $m\dot{\rm v}=-m_{\rm b}<\dot{\rm v}_{\rm y}>$ where $m_{\rm b}$ is 
the mass of a background particle, so that the drag force
is anti--parallel to the velocity of the perturber. In order to evaluate the
integrals over the wave numbers in eq.~(\ref{eq16}) it is advantageous to switch
from Cartesian form ${\rm k}_{\rm x}, {\rm k}_{\rm y}, {\rm k}_{\rm z}$ to
a mixed representation ${\rm k}_{\rm y}$, 
$k = \sqrt{{\rm k}_{\rm x}^2+{\rm k}_{\rm y}^2+{\rm k}_{\rm z}^2}$,
$\rmn{arctan}({\rm k}_{\rm x}/{\rm k}_{\rm z})$, and we obtain for the
deceleration the general result
\begin{eqnarray}
\dot{\rm v}=-\frac{4\pi G^2 m m_{\rm b} n_{\rm b}}{{\rm v}_0^2}
\left[ \rmn{erf}
\left( \frac{{\rm v}_0}{\sqrt{2}\sigma}\right)-\sqrt{\frac{2}{\pi}}
\frac{{\rm v}_0}{\sigma}e^{-\frac{{\rm v}_0^2}{2\sigma^2}}\right] \, 
\rmn{ln}\Lambda\,,
\label{eq18}
\end{eqnarray}
where erf denotes the usual error function. The Coulomb logarithm is defined as
\begin{equation}
\rmn{ln}\Lambda = \frac{4\pi^4}{G^2m^2}\int^{k_{\rm max}}_{k_{\rm min}}
dk\ k^3|\Phi_{\rm k}|^2
\label{eq19}
\end{equation}
In the case of a point mass formula (\ref{eq7}) implies 
$\Lambda=k_{\rm max}/k_{\rm min}$.
This result was first obtained in this form by \citet{K72} and is identical 
to Chandrasekhar's formula (\citealt{C43}), if $m+m_b\approx m$. The Coulomb
logarithm diverges in the familiar way both on small and large scales, i.e.~at
$k_{\rm max}^{-1}$ and $k_{\rm min}^{-1}$, respectively.

The Coulomb logarithm of the dynamical friction force exerted on a Plummer
sphere can be calculated by inserting eq.~(\ref{eq9}) into (\ref{eq19}) 
leading to
\begin{equation}
\rmn{ln}\Lambda = r_0^2\int^{\infty}_{k_{\rm min}}dk\ k\, K^2_1(r_0k)\,.
\label{eq20}
\end{equation}
If the Plummer radius $r_0$ shrinks to zero, expression (\ref{eq20}) changes
smoothly into the Coulomb logarithm of a point mass, because 
$\lim_{r_0\to 0}\ r_0K_1(k_0k)=k^{-1}$. The integral over the square of the
Bessel functions in eq.~(\ref{eq20}) can be evaluated using formula (5.54) of 
\citet{GR00},
\begin{equation}
\rmn{ln}\Lambda = -\frac{ r_0^2k_{\rm min}^2}{2} \left[ K^2_1(r_0k_{\rm min})
-K_0(r_0k_{\rm min})K_2(r_0k_{\rm min})\right]
\label{eq21}
\end{equation}
which is approximately
\begin{equation}
\rmn{ln}\Lambda  \approx  -\frac{1}{2}-\rmn{ln}(r_0k_{\rm min})\,,
\label{eq21a}
\end{equation}
in the limit of $ r_0k_{\rm min}\ll 1$.
This modified Coulomb logarithm converges on small scales precisely as 
found by \citet{W76}, but still diverges on large scales. A natural cut--off 
will be then the size of the stellar system under consideration.

For a perturber with the density distribution of a Hernquist profile we find
a Coulomb logarithm of the form
\begin{eqnarray}
\rmn{ln}\Lambda=\int^{\infty}_{k_{\rm min}}dk\ \frac{1}{k}[1 + r_0k\,
\rmn{cos}(r_0k)\rmn{si}(r_0k) \nonumber\\
- r_0k\,\rmn{sin}(r_0k)\rmn{ci}(r_0k)]^2\,.
\label{eq22}
\end{eqnarray}
It can be shown using the asymptotic expansions of the sine-- and 
cosine--integrals given by \citet{AS72} that the integrand in expression 
(\ref{eq22}) falls off at large $k$ as $4r_0(r_0k)^{-5}$. Thus the Coulomb 
logarithm converges at small scales. This is expected because, although the 
density distribution has an inner density cusp, the deflecting mass `seen' 
by a field star with a small
impact parameter scales with square of the impact parameter. At small wave 
numbers a Taylor expansion shows that the square bracket in expression
(\ref{eq22}) approaches 1 so that we find a logarithmic divergence of the 
Coulomb logarithm as in the case of the Plummer sphere.

\begin{figure}
\centering
\resizebox{\hsize}{!}{\rotatebox{0}{\includegraphics{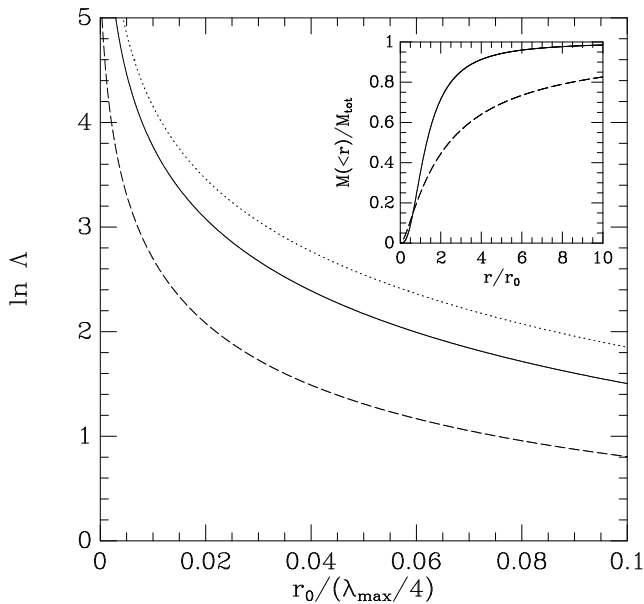}}}
\caption{Coulomb logarithms of the Plummer sphere (solid line), and a sphere 
with a Hernquist density profile (dashed line). The dotted line indicates
$-{\rm ln}(2 \pi r_0/\lambda_{\rm max})$. $r_0$ denotes the radial scale lengths
of the spheres and $\lambda_{\rm max}$ is the upper cut--off of the wavelength 
of the density perturbations (see text). The inlet shows the cumulative mass 
distributions of the Plummer and Hernquist models.}
\label{fig1}
\end{figure}

In Fig.~1 we illustrate the Coulomb logarithms of the Plummer sphere and a
sphere with a Hernquist density distribution according to eqns.~(\ref{eq21}) 
and (\ref{eq22}) as function of $r_0k_{\rm min}$.
Fig.~1 shows clearly that at given mass small sized perturbers experience 
a stronger dynamical friction force than larger ones. Since the cut-off of the 
wave number at $k_{\rm min}$ is determined by the radial extent of the stellar
system, which corresponds roughly to one half of the largest subtended wave 
length $\lambda_{\rm max}=2\pi /k_{\rm min}$, we use actually 
$r_0/(\lambda_{\rm max}/4)$ as abscissa in Fig.~1. For comparison we have also 
drawn $- ln(r_0k_{\rm min})$ in Fig.~1. The insert shows the cumulative mass 
distributions of both mass models. The half mass radius of the 
Hernquist model measured in units of $r_0$ is about twice as that of the 
Plummer sphere.
Thus for a proper comparison of the drag forces exerted on a Plummer sphere 
and a sphere with a Hernquist density profile the dashed line in Fig.~1 should 
be stretched by a factor of about 2 towards the right. But it is clear from 
Fig.~1 that the drag force exerted on a Plummer is always larger that the drag 
on a sphere with a Hernquist density profile. This is to be expected because of 
its shallower density profile. The logarithm $- ln(r_0k_{\rm min})$, although 
being the asymptotic expansion of the Coulomb logarithms (\ref{eq21}) and 
(\ref{eq22}) for $r_0k_{\rm min} \rightarrow 0$, is not a good approximation at 
larger  $r_0k_{\rm min}$. There is a systematic off--set relative to 
the true Coulomb logarithms which is given explicitely in eq.~(\ref{eq21a}) for 
the case of the Plummer sphere.

\begin{figure}
\centering
\resizebox{\hsize}{!}{\rotatebox{0}{\includegraphics{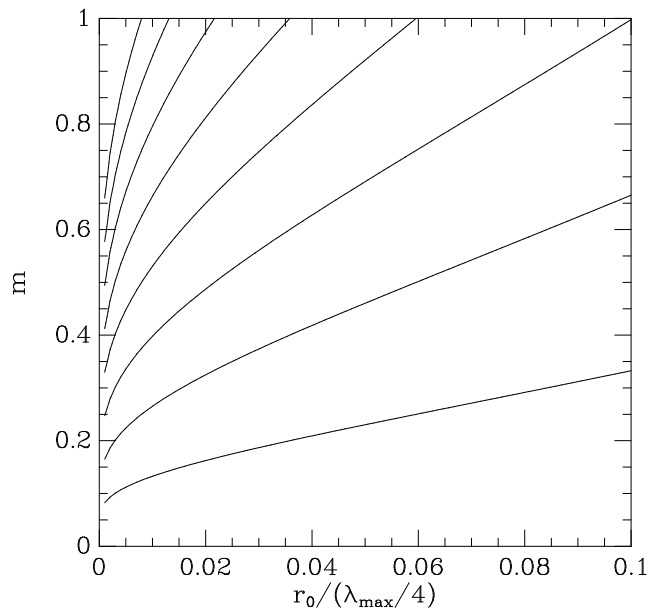}}}
\caption{Contour plot of $m\cdot{\rm ln} \Lambda$ for a Plummer sphere. The mass
$m$ is given in arbitrary units. The contour levels span the rnage from 0.5 to
4 at equidistant intervals of 0.5.}
\label{fig2}
\end{figure} 
So far we have treated the moving perturbers as rigid bodies. In reality, 
perturbers can be deformed by tidal fields. One source of the tidal field is the
induced polarization cloud itself. However, its effect is expected to be small. 
The moving perturber induces the polarization cloud in the background medium 
which reacts back on the perturber. 
Thus the dynamical friction force and any tidal fields exerted by 
the polarization cloud is of the order $G^2$ (cf. eq.~\ref{eq18}). The 
deformation of the perturber can be viewed as mass excesses and deficiencies. 
The momentum imparted by these `extra' masses to the particles of the 
background medium will be of the order of $G^3$ and can be safely considered
as a higher order effect.

There is a further effect if the perturbers are gravitationally bound systems
themselves like globular clusters or dwarf satellite galaxies. 
Such objects can and do loose mass due
to tidal shocking. This mass loss is primarily driven by tidal shocking due
to the shrinking of the tidal radii at the inner pericenters of the orbits of
the perturbing objects or by disc shocking when they pass through the galactic
disc (\citealt{GO97}, \citealt{GHO99}). Many studies, which cannot be all
enumerated here, have shown that mass loss,
which reduces dynamical friction, plays thus an important role for the evolution
of the orbits of the satellite galaxies. In order to illustrate the effect of
mass loss, on one hand, and the effect of a finite size of the perturber, on
the other hand, we show in Fig.~2 the variation of the deceleration by 
dynamical friction exerted on a Plummer sphere as function of the mass $m$ and 
the radial scale length $r_0$. As can be seen from the contour
plot in Fig.~2 both effects can be of comparable magnitude. If one considers, 
for example, a perturber with a scale length of 0.01 $(\lambda_{\rm max}/4)$, 
a doubling of its size has the same effect as a mass loss of 18 percent of the
original mass.

Finally we note that our analysis can be extended in a straightforward way, to
anisotropic velocity distributions of the field stars. \citealt{FA05} have 
shown that, if the velocity ellipsoid is either prolate or oblate, the velocity
dispersion in the solution of the Boltzmann equation (\ref{eq5}) is replaced by 
an effective velocity dispersion which depends on the semi-axes of the velocity 
ellipsoid and its orientation relative to the wave vector ${\bf k}$. This 
complicates the evaluation of the integrals with respect to the wave numbers 
in eq.~(\ref{eq16}) considerably. We hope to address this problem in a 
forthcoming paper (Esquivel \& Fuchs, in preparation).

\section{Acknowledgements}
We thank Andreas Just for useful discussions. Thanks are also due to the
anonymous referee for helpful comments. O.E.~gratefully acknowledges
financial support by the International-Max-Planck-Research-School for Astronomy
and Cosmic Physics at the University of Heidelberg. 

{}

\end{document}